\documentstyle[preprint,tighten,aps,amssymb]{revtex} 

\title{Gauge Transformations in Quantum Mechanics and 
the Unification of Nonlinear Schr\"odinger Equations} 

\author{H.-D.~Doebner\thanks{Electronic-mail address: 
ashdd@pt.tu-clausthal.de}}
\address{Arnold Sommerfeld Institute for Mathematical Physics and 
Institute for Theoretical Physics (A),
Technical University of Clausthal, D-38678 Clausthal-Zellerfeld, 
Germany}
\author{G.\,A.~Goldin\thanks{Electronic-mail address: 
gagoldin@dimacs.rutgers.edu}}
\address{Departments of Mathematics and Physics, 
Rutgers~University,
New~Brunswick,~New~Jersey~08903, USA}
\author{P.~Nattermann\thanks{Electronic-mail address: 
aspn@pt.tu-clausthal.de}}
\address{Institute for Theoretical Physics (A), 
Technical University of Clausthal,
D-38678~Clausthal-Zellerfeld, Germany}
\preprint{ASI-TPA/21/96}

\let\ds\displaystyle
\let\mathbbm\bf
\let\mathscr\mathcal

\def\rz{{\mathbbm R}}

\def\gz{{\mathbbm Z}}

\def\x{{\bf x}} 
\def\J{{\bf J}}
\def\v{\mbox{\boldmath{$\nu$}}}
\def\a{\mbox{\boldmath{$\alpha$}}}
\def\i{\mbox{\boldmath{$\iota$}}}
\def\Ref#1{(\ref{#1})}
\def\op#1{{\mathbf #1}} 
\def\A{{\bf A}}

\def\G{{\mathcal G}}

\def\H{{\mathscr H}}
\def\F{{\mathcal F}}

\def\N{{\mathcal N}}
\def\U{{\mathcal U}}
\def\R{{\mathcal R}}
\def\grad{\nabla}       
\def\dt{\partial_t}
\def\Aff{\mbox{\itshape Aff}}
\def\pp{\psi^{\,\prime}}
\def\np{\nu^\prime}
\def\mp{\mu^{\,\prime}}
\def\kp{\kappa^{\,\prime}}
\def\xp{\xi^{\,\prime}}
\def\ap{\alpha^{\,\prime}}

\def\Im{\mbox{Im}\,}
\def\stimes{\otimes_s}
\def\X{\mathrm{Vect}}
\begin{document}
\maketitle
\begin{abstract}
Beginning with ordinary
quantum mechanics for spinless
particles, together with the hypothesis
that all experimental measurements consist of positional measurements 
at different times, we characterize directly a class of
nonlinear quantum theories physically
equivalent to linear quantum mechanics through nonlinear gauge 
transformations.
%
%
We show that under
two physically-motivated assumptions,
these transformations are uniquely determined: they are exactly the 
group of
time-dependent, nonlinear gauge transformations introduced previously 
for a family of nonlinear Schr\"odinger equations.
The general equation in this family, including terms considered by 
Kostin, by
Bialynicki-Birula and Mycielski,
and by Doebner and Goldin,
with time-dependent coefficients,
can be obtained from the linear
Schr\"odinger equation through gauge transformation and a subsequent 
process we call {\it gauge generalization.\/} We thus unify, on 
fundamental grounds, a rather diverse set of nonlinear 
time-evolutions in quantum mechanics. 
\end{abstract}

\section{Introduction}\label{1}
Recently a group $\N$ of nonlinear gauge transformations was 
introduced
and shown to act as a transformation group in a family $\,{\F}\,$ of 
nonlinear Schr\"odinger equations (NLSEs). \cite{DG1996}
The family $\,{\F}\,$ consists of equations with nonlinear terms of 
the type introduced by Kostin \cite{Kos1972}, by Bialynicki-Birula 
and Mycielski \cite{BM1976}, by Guerra and Pusterla \cite{GP1982}, 
and by Doebner and Goldin \cite{DG1992,DG1994}, with time-dependent 
coefficients.

A transformation $N_{(\gamma, \Lambda)} \in \N$ is labeled by two 
real, time-dependent parameters $\gamma$ and $\Lambda$ (with $\Lambda 
\not= 0$), and
acts as a nonlinear analogue of a gauge transformation in quantum 
mechanics. Letting the time-dependent wave function $\psi(\x,t)$ on
$\rz^3$ be an arbitrary solution of any particular NLSE in $\,{\F}$,
$N_{(\gamma, \Lambda)}$ is given by
\begin{equation}
\psi^{\,\prime} = N_{(\gamma,\Lambda)}[\psi] = |\psi|\, 
\exp{[\,i(\gamma\ln|\psi|+\Lambda\arg\psi)\,]}. \label{nlt}
\end{equation}
Then $\psi^{\,\prime}$ solves a transformed equation that also 
belongs to $\,{\F}$.

The physical interpretation of this construction, developed briefly 
below,
was elaborated in some detail in Ref. \cite{DG1996}. However, the 
underlying mathematical
structure, and the physical reasons for the form of \Ref{nlt}, 
remained somewhat hidden.
Eq. \Ref{nlt} was motivated in earlier work by the desire to 
linearize the equations in a special subset of $\,{\F}$, and to 
obtain stationary and nonstationary solutions 
\cite{DG1994,N1994,N1995,DGN1995,G1995,NS1995}. The present paper 
takes a different, more fundamental approach to nonlinear gauge 
transformations and their consequences. 

We begin with linear, nonrelativistic quantum mechanics for spinless 
particles in $\rz^3$, together with the assumption, advocated for 
instance in
Refs. \cite{FH1965,N1966,M1974}, and discussed in 
Refs.~\cite{DG1996,DGN1995,G1995,L1995,N1997}, that all experimental 
measurements consist fundamentally of positional measurements made at 
different times. Defining as usual
the positional probability density
$\rho\,(\x,t) = \overline{\psi(\x,t)}\psi(\x,t)$, where $\psi$ 
conventionally is a normalized solution of the linear Schr\"odinger 
equation,
we are therefore interested in transformations $N$ which leave 
$\rho\,(\x,t)$
invariant --- i.e., such that for all $\psi$ in an appropriate domain 
of
the unit sphere in the Hilbert space $\H$, %
\begin{equation}
\label{rho_cons}
\overline{N[\psi](\x,t)}N[\psi](\x,t)=
\overline{\psi(\x,t)}\psi(\x,t)\,.
\end{equation}

In addition, $N$ should respect the prescription for writing the wave 
function subsequent to an ideal positional measurement. A 
conventional prescription for such a measurement at time $t_1$ 
consists of a projection in a region $B$ of position space (a Borel 
subset of $\rz^3$), with normalization, \begin{equation}
\label{psi_sub}
\psi_{s} (\x,t_1) = \left\{\begin{array}{c@{\quad}l} 
\frac{\psi(\x,t_1)}{\left(\int_B |\psi(\x,t_1)|^2 
d^3x\right)^{1/2}}\,, & \x\in B
\\
\\
0\,, & \x\not\in B
\end{array}\right.
\end{equation}
followed by time-evolution of $\psi_{s}(\x,t)$ for $t>t_1$ (here the 
subscript ``s'' stands for ``subsequent''). As $N$ should respect 
this prescription, we need for all $\psi$, $\x$
and $t\geq t_1$,
\begin{equation}
\label{rho_sub}
\big| N[\psi]_{s}(\x,t) \big|^2
= \big|\psi_{s}(\x,t)\big|^2\,,
\end{equation}
and because of \Ref{rho_cons},
\begin{equation}
\label{rho_sub2}
\big| N[\psi]_{s}(\x,t) \big|^2 = \big|N[\psi_{s}](\x,t) \big|^2\,. 
\end{equation}

We remark that in writing \Ref{psi_sub} we do not intend to express a 
commitment to a particular formalism for describing measurement. We 
merely note that the justification of $N$ as as a gauge transformation
requires that in addition to \Ref{rho_cons} it leave invariant the 
outcomes of
%
\emph{sequences} of positional measurements at various times. 
Eq.~\Ref{psi_sub} is one prescription for predicting such outcomes in 
quantum mechanics.

Now if all actual measurements (outcomes of experiments) are obtained 
from positional measurements performed at various times, it can be 
argued that a system with states $\psi$ obeying the Schr\"odinger 
equation, and one with states $N[\psi]$ obeying a transformed
equation, have the same physical content. But we make two essential 
observations:
\begin{itemize}
\item[(a)]
Eqs. \Ref{rho_cons} and \Ref{rho_sub}-\Ref{rho_sub2} do not require 
$N$ to be a linear
transformation --- nonlinear $N$ are also possible. \item[(b)]
Such nonlinear choices of $N$ will
transform a system governed by the usual, linear Schr\"odinger 
equation to physically equivalent systems obeying NLSEs that are, of 
course, linearizable (by construction).
\end{itemize}

The usual formulation and interpretation of quantum mechanics is 
based quite deeply on linearity and linear structures --- 
superposition principle, on observables modeled by self-adjoint
linear operators, on
a linear time-evolution equation for
the states, on a measurement process
involving orthogonal projection onto
linear subspaces for all sorts of observables, and on the description 
of mixed
states by density matrices.
Any proposal for nonlinearity
in quantum mechanics requires a revised
mathematical formulation and physical interpretation of all these 
ideas.
Here the linearizable NLSEs obtained using $N$ can be useful. Due to 
their physical equivalence with linear quantum mechanics, they serve 
as a kind of ``laboratory'' for exploring
how to generalize quantum
mechanics to accommodate nonlinearities. 

When $N$ is {\it assumed\/} to be linear (and densely defined), 
Eq.~\Ref{rho_cons} implies that it is a {\it unitary
multiplication operator\/} for each $t$. Then $N$ is labeled by a 
measurable function $\theta(\x,t)$, and we have
\begin{equation}
\label{gauge2}
\pp(\x,t) \,=\, \left(\op{U}_\theta\,\psi\right)(\x,t) \,=\, 
\exp{[i\theta(\x,t)]}\,\psi(\x,t)\,. \end{equation}
Any such $\op{U}_\theta$ commutes with the projection in 
\Ref{psi_sub}, thus ensuring \Ref{rho_sub2} and respecting the 
conventional prescription for wave functions subsequent to a 
positional measurement. 

If $\theta$ is independent of $\x$ and $t$, we have just introduced a 
fixed phase, sometimes called a
``gauge transformation of the first kind.'' This changes neither the 
Schr\"odinger equation nor the form of position and momentum 
operators.
A space- and time-dependent, linear
$U(1)$-gauge transformation, implemented by \Ref{gauge2}, is sometimes
called a ``gauge transformation of
the second kind.'' Such transformations constitute an abelian group 
$\U_{\,loc}$ of \emph{local} unitary operators acting on ${\mathcal 
H}$.
The physical equivalence of the two theories, with states $\psi$ and 
$\psi^\prime$ respectively,
is guaranteed by the invariance of the outcomes of sequences of 
positional observations at all times.

A system with wave functions governed by the (linear) Schr\"odinger 
equation
\begin{equation}
\label{SE}
i\hbar \dt\psi = -\frac{\hbar^2}{2m}\nabla^2 \psi + V\, \psi 
\end{equation}
is transformed by \Ref{gauge2} to a physically equivalent system, 
with wave function $\pp$ and Schr\"odinger equation %
\begin{equation}
\label{SE-trans2}
i\hbar\dt\pp = \frac{1}{2m}\left(\frac{\hbar}{i}\nabla 
-\hbar\grad\theta\right)^2 \pp +\left(V -\hbar\dt\theta\right) \pp \,.
\end{equation}
This observation
suggests a \emph{generic way} to construct new systems that are {\em 
not\/} physically equivalent to the original family given by 
\Ref{SE}-\Ref{SE-trans2}. The scalar term $-\hbar\dt\theta$ and the 
vector term $\hbar\grad\theta$ are merely special choices. If we take 
replace $-\hbar\dt\theta$ by a \emph{general} scalar field 
$\hat{\Phi}(\x,t)$ and $\hbar\grad\theta$ by a \emph{general} vector 
field
$\hat{\A}(\x,t)$, calling them (abelian) {\em gauge 
fields\/}, we obtain two well-known and well-established structures: 
\begin{enumerate}
\item[(a)]
a family $\F_{(\hat{\Phi},\hat{\A})}$ of time-evolution equations 
labeled by the
gauge fields $\hat{\Phi}$ and $\hat{\A}$: %
\begin{equation}
\label{MSE}
i\hbar\dt\psi = \frac{1}{2m}\left(\frac{\hbar}{i}\nabla 
-\hat{\A}\right)^2 \psi + \left(V+\hat{\Phi}\right)\psi \,; 
\end{equation}
and
\item[(b)]
an action on this family by
the gauge transformations $\op{U}_{\theta}$, to establish
equivalence classes --- so that Schr\"odinger equations with 
$(\hat{\Phi},\hat{\A})$ and with 
$(\hat{\Phi}^{\,\prime},\hat{\A}^{\,\prime}) 
= (\hat{\Phi}-\hbar\dt\theta$, $\hat{\A}+\hbar\grad\theta)$ describe 
physically equivalent systems --- with the family being closed under 
the action of gauge transformations.
\end{enumerate}
This generic construction, which we here call \emph{gauge 
generalization}, is
physically relevant because external electromagnetic fields 
$(\Phi,\A)$ interacting with a charged particle provide a realization 
of $(\hat{\Phi},\hat{\A})$ in nature: in Gaussian units, 
$\hat{\Phi}=e\Phi$ and $\hat{\A}=(e/c)\A$, where $e$ is the charge of 
the particle.
The gauge-transformed Schr\"odinger equations are physically 
equivalent to the original,
but those obtained from them by gauge generalization are not. These 
well-known
results provide a model for similar arguments involving the nonlinear 
transformations $N$. 

In Section~\ref{2}, we demonstrate that two straightforward, 
physically-motivated conditions precisely specify the group $\N$ of 
time-dependent, nonlinear gauge transformations introduced in 
Ref.~\cite{DG1996}. These assumptions are: (a) strict locality and
(b) a separation condition. We observe that \Ref{rho_sub2} is then 
ensured.

In Section~\ref{3}, we
apply various subgroups of $\N$ to the
linear Schr\"odinger equation
\Ref{SE}. This leads to
physically equivalent systems satisfying NLSEs, where the 
coefficients obey certain constraints. Then, in structural analogy to 
the way
\Ref{SE-trans2} motivates \Ref{MSE}, we construct new, physically 
{\em inequivalent\/} systems by generalizing the parameters so as to 
break the constraints.

In Section~\ref{4}, following this analogy, we consider the 
parameters as {\em gauge parameters.\/}
We thus obtain a family of NLSEs through gauge generalization and 
gauge closure, labeled by the gauge parameters, on which the gauge 
group acts to establish physical equivalence classes.
In this way, we derive naturally --- as a unified class --- equations 
containing the terms proposed by Kostin, Bialynicki-Birula and 
Mycielski,
and Doebner and
Goldin, with coefficients that are (in general) time-dependent. The 
subfamily
that includes the equations of Guerra and Pusterla turns out to be 
equivalent to linear quantum mechanics.

of mixed states in nonlinear theories, 
physical assumptions. 

We believe this to offer a fundamentally new perspective, partially 
elucidating the hidden mathematical and physical structure behind 
certain nonlinear quantum time-evolutions.
\section{Conditions on nonlinear gauge transformations} \label{2} 
\subsection{Locality}\label{2.1}
We have from Eq.~\Ref{rho_cons} that
\begin{equation}
\label{Ntheta}
N[\psi](\x,t) = \exp [i G_\psi (\x,t)] \psi(\x,t)\,, \end{equation}
where $G_\psi$ is a real-valued function of $\x$ and $t$ depending on 
$\psi$. It
is apparent that $G_\psi$ must be further restricted if for instance 
we hope
to ensure \Ref{rho_sub2} for all Borel subsets $B$ of $\rz^3$. 
Suppose the value of $G_\psi(\x,t)$ at $\x=\x_1$ depends nontrivially 
on values of
$\psi(\x,t)$ for $\x\neq\x_1$ and the evolution equation is local. 
Then we will be unable to satisfy
\Ref{rho_sub}-\Ref{rho_sub2} in the general case of a region $B$ 
where $\x_1\in B$ but $\x\not\in B$. Therefore let us assume $N$ to 
be a {\em local\/} transformation, in analogy with the linear gauge 
transformations $\op{U}_\theta$. This is taken here in the strict 
sense that the value of $N[\psi]$ at $(\x,t)$ should depend only on 
$\x$, $t$, and the value of $\psi(\x,t)$ --- not
on any other space or time points, and not on derivatives of $\psi$. 
Then we must have
\begin{equation}
\label{local}
\pp (\x,t) = N_F\,[\psi](\x,t) = \exp\left[iF(\psi(\x,t),\x,t)\right] 
|\psi(\x,t)|\,,
\end{equation}
where $F$ is a real-valued function (defined up to integer multiples 
of $2\pi$) of the three variables whose values are provided by 
$\psi(\x,t)$, $\x$, and $t\,$ . The possible dependence of $F$ on the 
value of $\psi(\x,t)$ allows nonlinearity in $N_F$. With 
$R(\x,t)=|\psi(\x,t)|$ and $S(\x,t)=\arg \psi(\x,t)$, we can consider 
$F$ to be a function
of the real variables $R, S, \x$, and $t$, relaxing for now the 
requirement that $F$ take the same value at $S$ and $S + 2\pi n$.

Note that a weaker assumption, in which $F$ is permitted to depend on 
finitely many derivatives of $\psi$ at $(\x,t)$, may still be 
compatible with \Ref{rho_sub2}. We make a stricter assumption here, 
which
limits the resulting time-evolution
equations to second order.

\subsection{A separation condition}\label{2.3} We consider now 
systems of $n$ particles described by normalized states in
$\H^{(n)} = L^2(\rz^{3n},d^{\,3n}\!x)$.
For simplicity, take each individual particle to evolve under the same
time-evolution operator $T^{(1)}$.
We suppose a hierarchy of
time-evolutions $T^{(n)}$ of $n$-particle states, fulfilling the 
\emph{separation condition}.
For linear time evolutions 
this condition requires that
product states $\psi^{(n)} =
\psi_1\otimes\ldots\otimes\psi_n$, $\|\psi_j\|=1$, $j=1,\ldots,n$, 
evolve into product states: %
\begin{equation}
\label{T-sep}
T^{(n)}\left[\psi^{(n)}\right] = T^{(1)}[\psi_1]\otimes 
T^{(1)}[\psi_2] \otimes\ldots \otimes T^{(1)}[\psi_n]\,. 
\end{equation}
It ensures that in the absence of interaction terms, initially 
uncorrelated subsystems remain uncorrelated, and $T^{(n)}$ is 
extended (by linearity) from product states to all of $\H^{(n)}$. 

It is physically plausible to assume \Ref{T-sep} for nonlinear time 
evolutions $T^{(n)}$ as well \cite{BM1976,GSv1994}. Then nonlinear 
gauge transformations $N_F^{(n)}$ should respect this condition. Here 
Eq.~\Ref{local},
the states $\pp_j(\x,t)=N_{F}[\psi_j]$ in \Ref{local} are governed by 
a nonlinear time evolution ${T^{(1)}}^\prime$, and our separation 
condition becomes for product states ${\psi^{(n)}}^\prime 
= \pp_1\otimes\ldots\otimes\pp_n$, $\|\pp_j\|=1$, $j=1,\ldots,n$,
\begin{equation}
\label{Tp_sep}
{T^{(n)}}^\prime\left[{\psi^{(n)}}^\prime\right] = 
{T^{(1)}}^\prime[\psi_1^\prime]\otimes \cdots \otimes 
{T^{(1)}}^\prime[\psi_n^\prime]\,.
\end{equation}
We thus want a nonlinear gauge transformation $N^{(n)}_{F}$ acting on 
the unit sphere in $\H^{(n)}$, with $\|N_F^{(n)}[\psi]\|=1$, so that 
on the product states $\psi^{(n)}$ \begin{equation}
\label{N-sep}
N^{(n)}_{F}[\psi] =
N_{F}[\psi_1] \otimes\ldots\otimes
N_{F}[\psi_n]\,.
\end{equation}
Unitary gauge transformations $\op{U}^{(n)}$ in $\H^{(n)}$ may be 
written as
\begin{equation}
\label{Un}
\left(\op{U}\psi^{(n)}\right)(\x_1,\ldots,\x_n,t) = 
\exp[i\theta_n(\x_1,\ldots,\x_n,t)] \psi(\x_1,\ldots,\x_n,t)\,. 
\end{equation}
On product states, using \Ref{gauge2},
we want
\begin{equation}
\label{theta-sep}
\op{U}^{(n)}\psi^{(n)} = (\op{U}_\theta\psi_1)\otimes 
(\op{U}_\theta\psi_2)
\otimes\ldots \otimes (\op{U}_\theta\psi_n) \,, \end{equation}
so that
\begin{equation}
\label{theta-sum}
\theta_n(\x_1,\x_2,\ldots,\x_n,t) = \sum_{j=1}^n \theta(\x_j,t). 
\end{equation}
And of course, for this case, the operators are linear and can be 
extended by linearity from product states to the whole Hilbert space. 
But $N^{(n)}_{F}$ in Eq.~\Ref{N-sep} is nonlinear, so we cannot 
extend it uniquely to $\H^{(n)}$. 

The apparently weak condition \Ref{N-sep} leads nevertheless to a 
sharp restriction on $F$. To see this it is sufficient to discuss the 
case $n=2$. With $R_j=|\psi_j|$, $S_j=\arg\psi_j$, $j=1,2$, 
Eqs.~\Ref{local} and \Ref{N-sep} imply that \begin{equation}
\label{N-sep2}
F(R_1,S_1,\x_1,t) + F(R_2,S_2,\x_2,t)
\end{equation}depends only on the product $R=R_1R_2$ and the sum 
$S=S_1+S_2$. Thus \begin{equation}
F(R_1,S_1,\x_1,t) + F(R_2,S_2,\x_2,t) = F(R,S,\x_1,t) + 
F(1,0,\x_2,t)\,,
\end{equation}
for all $R,S,\x_1,\x_2,t$, whence $F(R_2,S_2,\x_2,t) - F(1,0,\x_2,t)$ 
must be independent of $\x_2$. Setting $F(1,0,\x,t) = \theta(\x,t)$ 
and $L(R,S,t) = F(R,S,\x,t)-\theta(\x,t)$, we have the functional 
equation \begin{equation}
\label{Leq}
L(R_1,S_1,t) + L(R_2,S_2,t) = L(R_1R_2,S_1+S_2,t)\,. \end{equation}
The smooth solutions of \Ref{Leq} are given by $L(R,S,t)= \gamma(t) 
\ln R +\Lambda(t)S$, where $\gamma,\Lambda$ are real functions of 
$t$. Non-degeneracy of the transformation requires $\Lambda(t)\neq 
0$. Finally, we have \begin{equation}
\label{FgLt}
F(R,S,\x,t) = \gamma(t) \ln R + \Lambda(t) S + \theta(\x,t)\,. 
\end{equation}
The above argument is similar to the way in which generalized 
homogeneity of the time-evolution is deduced from the separation 
property \cite{GSv1994}.

Note that our separation condition is \emph{weak} in the sense that 
for nonlinear $T^{(n)}$ and $N_F^{(n)}$ it is only defined on product 
states; for non-product (entangled) initial states, non-interacting 
subsystems may
yet acquire new correlations.
The nonlocal effects in some nonlinear evolution equations can be 
traced back to this fact \cite{LN1997}. A \emph{strong} version of 
the separation condition, more adapted to the
physical situation and valid for general states, can be formulated 
along the lines given in Ref. \cite{LN1997}. 

\subsection{The result}\label{2.4}

In short, the locality and the separation condition required on 
$n$-particle product states boils down the transformations $N_{F}$ 
for single particle states to those labeled by two real functions 
$\gamma$ and
$\Lambda$ of time, with $\Lambda$ non-vanishing, and a real function 
$\theta$ of space and time:
\begin{equation}
\label{NgLt}
\left(N_{(\gamma,\Lambda,\theta)}[\psi]\right)(\x,t) = |\psi(\x,t)| 
\exp\left[i\left(\gamma(t)\ln |\psi(\x,t)| +\Lambda(t)\arg\psi(\x,t) 
+\theta(\x,t)\right)\right]\,.
\end{equation}
The set $N_{(\gamma,\Lambda,\theta)}$ forms a group $\G$, with 
multiplication law
\begin{equation}
\label{glt-mul}
N_{(\gamma',\Lambda',\theta')} \circ N_{(\gamma,\Lambda,\theta)} 
=N_{(\gamma'+\Lambda'\gamma,\Lambda'\Lambda,\theta'+\Lambda'\theta)}\,. 
\end{equation}
This can be expressed in terms of $3\times 3$ matrices, 
\begin{equation}
\label{3x3mat}
N_{(\gamma,\Lambda,\theta)} \simeq \left(\begin{array}{ccc} 
1 & 0 & 0 \\
\theta & \Lambda & 0\\
\gamma & 0 & \Lambda
\end{array}\right)
\end{equation}
with entries $\Lambda=\Lambda(t)$, $\gamma=\gamma(t)$, and 
$\theta=\theta(\x,t)$ taken from the corresponding function spaces. 
We thus have here a group $\G$ of nonlinear gauge transformations, 
strictly local and separating on $n$-particle product states, labeled 
by time-dependent parameters $\gamma$ and $\Lambda$ together with a 
function $\theta(\x,t)$.
The group is a semi-direct product of the group of gauge 
transformations of the
second kind $\U_{\,loc}=\{\op{U}_\theta\}$ and the group
$\N$, mentioned in the introduction, of `pure nonlinear' gauge 
transformations (where $\theta\equiv 0$): \begin{equation}
\G = \N \stimes \U_{\,loc}\,.
\end{equation}
$\G$ can be viewed as a {\em
nonlinear generalization\/} of $\U_{\,loc}$, the group of `gauge 
transformations of the {\em third kind\/}' \cite{DG1996}.

The transformations $N_{(\gamma,\Lambda,\theta)}$ are not uniquely 
defined on the Hilbert space.
If we restrict the range of
$\Lambda$ to the integers, $\Lambda(t)\in\gz$, then 
$N_{(\gamma,\Lambda,\theta)}$ is well defined. Then if $\Lambda$ is a 
continuous function of time, $\Lambda$ has to be a constant; 
$N_{(\gamma,\Lambda,\theta)}$ is invertible with this restriction only
for $\Lambda=\pm 1$.
$\Lambda = -1$ corresponds to complex
conjugation: $N_{(0,-1,0)}\psi = \bar\psi$. $N_{(\gamma,1,\theta)}$ 
is strongly continuous \cite{L1995}, and the set of these 
transformations is an Abelian subgroup of $\G$, \begin{equation}
\G\supset \G_0 = \N_0 \otimes \U_{\,loc}\,. \end{equation}
where $\N_0:=\{N_\gamma:=N_{(\gamma,\Lambda\equiv 1,\theta\equiv 
0)}\}$.

For non-integer $\Lambda$, $N_{(\gamma,\Lambda,\theta)}$ may be 
specified uniquely on certain domains in the Hilbert space, e.g.\ by 
imposing continuity of the phase of $N(\psi)$ on a domain of 
non-vanishing functions $\psi$.
However, such a domain is not needed explicitly for our further 
considerations.
\subsection{A generalization}
Because of the difficulties with the separation condition mentioned 
above, a more general group structure is also of interest. This can 
be obtained. without assuming separation, by making a 
physically-motivated, weaker assumption: an intertwining relation 
that follows from requiring compatibility with linear gauge 
transformations.

The group of linear gauge transformations $\U_{\,loc}$ is 
commutative, but this need not be the case for the set $\{N_F\} 
\supset \U_{\,loc}$. In
particular, $\op{U}_\theta$ might not commute with $N_F$. We explore 
the condition that $N_F$ be consistent with the usual notion
of physical equivalence under gauge transformations of the second 
kind. That is, the result of applying $N_F$ to a gauge-transformed 
theory with wave functions $\op{U}_\theta\,\psi $ should be 
expressible as a transform by $\op{U}_{\theta^{\,\prime}}$ of the 
theory with wave functions $N_F[\psi]$, where, in general, 
$\theta^{\,\prime} (\x,t)\not= \theta(\x,t)$. Thus we
require an {\em intertwining relation\/} %
\begin{equation}
\label{intertwine}
N_F[\op{U}_\theta\,\psi ]= \op{U}_{\theta^{\,\prime}} N_F[\psi]\,.
\end{equation}
Here the function $\theta^{\,\prime}(\x,t)$ depends on both of the 
functions $F$ and
$\theta$.

Then Eq.~\Ref{intertwine} implies the functional equation %
\begin{equation}
\label{expf}
\exp \,i[F(R,S+\theta\,;\x,t)] = \exp \,i[\theta^{\,\prime}(\x,t) + 
F(R,S;\x,t)]\,,
\end{equation}
valid for each $R,\,S,\,\x$, and $t$.
It is straightforward to show that smooth solutions $F$ of \Ref{expf} 
take the form
$F(R,S; \x,t) = k(R,\x,t) + \lambda(\x,t)S$, where $k$ and $\lambda$ 
are real-valued functions of the indicated variables. Non-degeneracy 
of the transformation requires $\lambda(\x,t)\neq 0$ for all $\x,t$. 
Thus $N_F$ is parameterized by $k$ and $\lambda$, and given by %
\begin{equation}
\label{Nkl}
N_{(k,\lambda)}[\psi](\x,t) =
\exp \,i[k(|\psi(\x,t)|,\x,t) +
\lambda(\x,t) \arg \psi(\x,t)]\,\,|\psi(\x,t)|\,. \end{equation}
One easily checks that \Ref{rho_cons}, \Ref{local}, and 
\Ref{intertwine} are
fulfilled, with
\begin{equation}
\label{thetat}
\theta^\prime(\x,t)= \lambda(\x,t)\theta(\x,t)\,. \end{equation}

The set $\{N_{(k,\lambda)};\,\lambda(\x,t)\neq 0\}$ is a 
non-commutative, infinite dimensional group $\tilde\G$ with 
multiplication law
\begin{equation}
\label{kl-mul}
N_{(k,\lambda)}\circ N_{(k',\lambda')} = 
N_{(k+k'\lambda,\lambda\lambda')}\,.
\end{equation}
$N_{(0,1)}$ acts as the identity on
$\psi$, and $N_{(-k/\lambda,\,1/\lambda)}$ is the (formal) inverse of 
$N_{(k,\lambda)}$.
The group law may be expressed as multiplication of $2\times 2$ 
matrices
\begin{equation}
\label{Naff1}
N_{(k,\lambda)} \simeq \left(
\begin{array}{cc}
1 & 0 \\
k & \lambda
\end{array}\right)
\end{equation}
with entries $k(|\psi|,\x,t)$ and
$\lambda(\x,t)$ taken from function spaces. Such matrices span a 
linear representation $\Aff(1)$ of the one-dimensional affine group. 

The nonlinear transformations $N_{(\gamma,\Lambda,\theta)}$ are 
special cases of $N_{(k,\lambda)}$; i.e., the separation condition 
restricts $k$ and $\lambda$ to the form
\begin{eqnarray}
k(|\psi|,\x,t) &=& \gamma(t) \ln|\psi| + \theta(\x,t)\,, \\ 
\lambda(\x,t) &=& \Lambda(t) \,;
\end{eqnarray}
and $\G$ is a subgroup of $\tilde\G$.

\section{Nonlinear quantum-mechanical evolution equations 
from gauge generalization}\label{3}
\subsection{Linearizable NLSEs}\label{sec3.1} In accordance with the 
discussion in Section~\ref{1}, we are now interested in the evolution 
equation of \begin{equation}
\label{trans}
\pp(\x,t) = N_{(\gamma,\Lambda,\theta)}[\psi](\x,t) \,, \end{equation}
when $\psi(\x,t)$ is a solution of a linear {Schr\"odinger} equation 
\begin{equation}
\label{F0}
i\dt\psi = \left(\nu_1 \Delta + \mu_0 V\right)\psi\,. \end{equation}
Let us regard \Ref{F0} as belonging to a parameterized family 
$\F_0(\nu_1,\mu_0)$, $\nu_1\neq0$, depending on the two real 
parameters $\nu_1,\mu_0$; in Eq.~\Ref{SE}, $\nu_1= - \hbar/{2m}$ and 
$\mu_0= 1/{\hbar}$.

Due to \Ref{intertwine} linear gauge transformations can be treated 
independently, and we shall here restrict ourselves to the case 
$\theta\equiv 0$. Applying the group $\N$ to $\F_0$, we obtain a 
family $\overline{\F_0}$ of NLSEs \begin{equation}
\label{F}
i\dt\pp = \left(\np_1(t) \Delta + \mp_0(t) V + F_{DG}^{(0)}[\pp] + 
F_{BM}[\pp] + F_{K}[\pp]\right)\pp \,,
\end{equation}
where
\begin{eqnarray}
\label{F-DG0a}
F_{DG}^{(0)}[\pp] &=&
\mp_1(t)\left(\nabla \left( \Im \left\{\frac{\nabla\pp}{\pp} \right\} 
\right) +\frac{i}{2}\frac{\Delta|\pp|^2}{|\pp|^2}\right) +2\kp(t) 
\frac{\Delta |\pp|}{|\pp|} \,,\\*[2pt] \label{F-BM}
F_{BM}[\pp] &=& \ap_1(t) \log |\pp|^2 \,,\\*[8pt] \label{F-K}
F_{K}[\pp] &=& \ap_2(t)\arg\pp \,.
\end{eqnarray}
The coefficients $\np_1,\mp_0,\mp_1,\kp,\ap_1,$ and $\ap_2$ are 
constrained, and depend on both $\nu_1,\mu_0$, and on 
$\Lambda(t),\gamma(t)$:
\begin{equation}
\label{F0-para}
\begin{array}{c}
\ds
\np_1(t) = \frac{1}{\Lambda(t)} \nu_1\,,\quad \mp_0(t) = 
\Lambda(t)\mu_0\,,\quad
\mp_1(t) = -\frac{\gamma(t)}{\Lambda(t)}\nu_1\,,\quad \kp(t) = 
\frac{\gamma(t)^2+\Lambda(t)^2-1}{2\Lambda(t)}\nu_1\,,\\ \ds \ap_1(t) 
= \gamma(t) \frac{\dot\Lambda(t)}{2\Lambda(t)} 
-\frac{1}{2}\dot\gamma(t)\,,\quad
\ap_2(t) = -\frac{\dot\Lambda(t)}{\Lambda(t)}\,. \end{array}
\end{equation}
This family $\overline{\F_{0}}$ is closed under $\N$; i.e., it is the 
{\em gauge closure} of
$\F_0(\nu_{1},\mu_{0})$ under the action of the group $\N$. It is, up 
to questions of domain mentioned above, linearizable. It depends on 
the independent quantities $\nu_1$, $\mu_0$, $\gamma(t)$ and 
$\Lambda(t)$. One could also write $\overline{\F_0}$ as
$\F(\nu_1,\mu_0,\mu_1,\kappa,\alpha_1,\alpha_2)$ labelled by 
time-dependent coefficients that are constrained. 

Note that if $\Lambda$ and $\gamma$ are independent of $t$, the 
coefficients are time-independent, and
$\alpha_1^\prime=\alpha_2^\prime=0$.

If we restrict $\N$ to the subgroup $\N_0$, then starting with 
$\F_0(\nu_1,\mu_0)$, we obtain a family $\overline{\F_0}^0$ closed 
under $\N_0$ and contained in $\overline{\F_0}$; here the indexed bar 
denotes the closure with respect to $\N_0$. The elements in 
$\overline{\F_0}^0$ are by construction linearizable NLSEs. The 
parameters are \begin{equation}
\label{R0-para}
\begin{array}{c}
\ds \np_1 = \nu_1\,,\quad
\mp_0 = \mu_0\,,\quad
\mp_1(t) = \gamma(t)\nu_1\,,\quad
\kp(t) = \frac{\gamma(t)^2}{2}\nu_1\,,\\ \ds \ap_1(t) = 
-\frac{1}{2}\dot\gamma(t)\,,\quad 
\ap_2(t) = 0\,.
\end{array}
\end{equation}
Now the term $F_K$ disappears, everything is well-defined, and 
$\np_1$ and $\mp_0$ are time-independent invariants. Strictly 
speaking, these NLSEs are \emph{defined} using the continuity and 
invertibility of $N_{(\gamma,1,0)}$. 

For later purposes we mention that $F_{DG}^{(0)}$ decomposes into 
independent nonlinear real functionals $R$ with the following 
properties:
$R[\psi]$ is {Euclid}ean invariant, complex homogeneous of degree 
zero and a
rational function of $\psi,\bar\psi$ with derivatives not higher than 
second order in the numerator only. There exist five functionals of 
this type (see \cite{DG1994}): \begin{equation}
\label{Rj}
\begin{array}{c}
\ds R_1[\psi] = \frac{\nabla\cdot\J}{\rho}\,, \quad 
R_2[\psi] = \frac{\Delta \rho}{\rho}\,,\\ \ds R_3[\psi] = 
\frac{\J^2}{\rho^2}\,,\quad 
R_4[\psi] = \frac{\J\cdot\nabla\rho}{\rho^2}\,,\quad R_5[\psi] = 
\frac{(\nabla\rho)^2}{\rho^2}\,, \end{array}
\end{equation}
where $\rho=\bar\psi\psi$ and
$\J= ({1}/{2i})\,\left(\bar\psi\nabla\psi-(\nabla\bar\psi) 
\psi\right)$ are the
probability density and current corresponding to $\psi$. With this 
notation $F_{DG}^{(0)}$ in \Ref{F-DG0a} is a complex linear 
combination:
\begin{equation}
\label{F-DG0}
F_{DG}^{(0)}[\psi] = \mu_{1}(t)\left(R_{1}[\psi] - 
R_{4}[\psi]\right)+ i\nu_{2}(t) R_2[\psi] + 
\kappa(t)(R_{2}[\psi]-\frac{1}{2}R_{5}[\psi])\,,\\*[2pt] 
\end{equation}
with
\begin{equation}
\label{nu2}
\nu_2(t) = -\frac{1}{2}\mu_1(t)\,.
\end{equation}
The term $R_{3}[\psi]$ will appear in the next section. 

\subsection{Generalizing linearizable NLSEs; gauge 
parameters}\label{sec3.2}
The nonlinear gauge transformations $N_{(\gamma,\Lambda)}$ generate 
special linearizable NLSEs; i.e., nonlinear PDEs with constrained 
coefficients, physically equivalent to linear Schr\"odinger equations.
Hence the situation is similar to the case of gauge transformations 
$\op{U}_{\theta}$ in Section~\ref{1}. It is possible to construct 
generically through \emph{gauge generalizations} and \emph{gauge 
closures} a sequence of {new families of evolution equations} 
physically \emph{in}equivalent to the linear Schr\"odinger equation. 
We obtain the sequence of these families in three steps: 
\paragraph*{\bfseries Step 1:}
We break the constraints \Ref{F0-para} in $\overline{\F_0}$ (gauge 
generalization);
i.e., we take the six constrained
coefficients $\np_{1},\mp_{0},\mp_{1},\kp,\ap_{1},\ap_2$ as 
independent functions of time. Thus we obtain a family 
$\F_1\left(\nu_1,\mu_0,\mu_1,\kappa,\alpha_1,\alpha_2\right)$ with 
six independent parameters. The gauge
transformations $\N$ are automorphisms of this family. That is, 
$\overline{\F_1} = \F_1$; the family is gauge closed. In the notation 
of Ref.~\cite{DG1996}, $\kappa= \mu_2 - \frac{1}{2} \nu_1$.
\paragraph*{\bfseries Step 2:}
We break the constraint \Ref{nu2} for $F_{DG}^0$ in $\F_1$ (gauge 
generalization),
\begin{equation}
\label{F-DG1}
F_{DG}^{(1)}[\psi] = i\nu_2(t)R_2[\psi]
+ \mu_{1}(t) (R_{1}[\psi] - R_{4}[\psi]) + \kappa(t) 
(R_{2}[\psi]-\frac{1}{2}R_{5}[\psi])\,, \end{equation}
and obtain a seven-parameter family
$\F_2(\nu_1,\nu_2,\mu_0,\mu_1,\kappa,\alpha_1,\alpha_2)$ of NLSEs 
\Ref{F}, with $F_{DG}^{(1)}$ replacing $F_{DG}^{(0)}$. 

The action of the group $\N$,
however, does not leave this family invariant. The gauge closure 
$\overline{\F_2}$ of $\F_2$ consists of NLSEs \Ref{F} with 
\begin{equation}
\label{F-DG2}
F_{DG}^{(2)}[\psi] = i\np_2(t)R_2[\psi]
+ \mp_{1}(t)(R_{1}[\psi] - R_{4}[\psi])
+ \kp(t) R_{2}[\psi]
+ \xp(t) R_{5}[\psi]
\end{equation}
in place of $F_{DG}^{(1)}$.
Now there are eight coefficients. The next step is again to break any 
constraints, but the coefficients are already not constrained. Thus 
we write $\overline{\F_2}$ as a family 
$\F_3(\nu_1,\nu_2,\mu_0,\mu_1,\kappa,\xi,\alpha_1,\alpha_2)$ with 
eight time-dependent parameters, and
invariant by construction under
$\N$; i.e., $\overline{\F_2} = \F_3 = \overline{\F_3}$. In the 
notation of ref.~\cite{DG1996},
$\xi=\mu_5+\frac{1}{4}\nu_1$.
The explicit formula for these coefficients is given by 
Eq.~\Ref{F4c-para} below, with $\mu_3=-\nu_1$ and 
$\xi=-\frac{1}{2}\kappa$. \paragraph*{\bfseries Step 3:}
We write $\Delta\psi$ as a complex linear combination of 
$R_j[\psi]\,\psi$,
\begin{equation}
\label{Delta-exp}
\Delta\psi = \left(iR_1[\psi]+\frac{1}{2}R_2[\psi] -R_3[\psi] 
-\frac{1}{4}R_5[\psi] \right)\psi\,,
\end{equation}
insert into \Ref{F} and obtain an additional term $(\mu_3+\nu_1) 
R_3[\psi]$ in $F_{DG}$, and a constraint $\mu_3(t)=-\nu_1(t)$. 

We break this constraint, and obtain from $\F_3$ a family 
$\F_4(\nu_1,\nu_2,\mu_0,\mu_1,\kappa,\mu_3,\xi,\alpha_1,\alpha_2)$ 
depending on nine time-dependent parameters. 

The closure $\overline{\F_4}$ is larger than $\F_4$ and contains all 
NLSEs \Ref{F} with
\begin{equation}
\label{F-DG}
F_{DG}[\psi] = i\np_2 R_2[\psi]
+ \mp_1 R_1[\psi]
+ \kp R_2 [\psi]
+ (\mp_3+\np_1) R_3[\psi]
+ \mp_4 R_4 [\psi]
+ \xi R_5[\psi]\,,
\end{equation}
where the time-dependent coefficients are given by: 
\begin{equation}
\label{F4c-para}
\begin{array}{c}
\np_1 = \frac{\nu_1}{\Lambda}\,,\quad
\np_2 = -\frac{\gamma}{2\Lambda}\nu_1 +\nu_2\,,\quad 
\Lambda \mu_0\,,\quad
\mp_1 = -\frac{\gamma}{\Lambda}\nu_1 + \mu_1\,,\\[1ex] \kp = 
\frac{\gamma^2+\Lambda^2-1}{2\Lambda}\nu_1 
-\gamma\nu_2 -\frac{\gamma}{2}\mu_1 +\Lambda\kappa \,,\quad 
\mp_3 = \frac{1}{\Lambda}\mu_3\,,\quad
\mp_4 = \frac{\gamma}{\Lambda}\nu_1 - \mu_1 
-\frac{\gamma}{\Lambda}\mu_3\,,\quad \\[1ex] \xp = 
\frac{1-\gamma^2-\Lambda^2}{4\Lambda}\nu_1 
+\frac{\gamma}{2}\mu_1
+\frac{\gamma^2}{4\Lambda}\mu_3
+\Lambda\xi\,,\quad\\[1ex]
\ap_1 = \Lambda \alpha_1 -\frac{\gamma}{2}\alpha_2 
+ \gamma\frac{\dot\Lambda}{2\Lambda}
-\frac{1}{2}\dot\gamma\,,\quad 
\ap_2 = \alpha_{2}-\frac{\dot\Lambda}{\Lambda}\,. \end{array}
\end{equation}
These coefficients are actually independent, so that 
$\overline{\F_4}$ is a ten-parameter family. For a more symmetrical 
notation, we now go over to using $\mu_2=\kappa+\frac{1}{2}\nu_1$ and 
$\mu_5=\xi-\frac{1}{4}\nu_1$, denoting the family by 
$\F_5(\nu_1,\nu_2,\mu_0,\ldots,\mu_5,\alpha_1,\alpha_2)$: 
\begin{equation}
\label{NLSE}
\begin{array}{rcl}
\ds i \dt\psi &=&\ds i\sum_{j=1}^2 \nu_j R_j[\psi]\,\psi 
+ \sum_{k=1}^5 \mu_k R_k[\psi]\, \psi
+ \mu_0 V \psi
+ \alpha_1 \log|\psi|^2 \psi
+ \alpha_2 (\arg\psi) \psi\,,
\end{array}
\end{equation}
or in a form which exhibits the linear part separately, with 
Laplacian $\Delta$,
\begin{equation}
\label{NLSE2}
\begin{array}{rcl}
i \dt\psi&=&\ds \left(\nu_1\Delta + \mu_0 V\right)\psi 
+ i\nu_2R_2[\psi]\,\psi \\
&&      + \mu_1R_1[\psi]\,\psi
+ (\mu_2-\frac{1}{2}\nu_1)R_2[\psi]\,\psi + 
(\mu_3+\nu_1)R_3[\psi]\,\psi\\
&&      + \mu_4 R_4[\psi]\,\psi
+ (\mu_5+\frac{1}{4}\nu_1) R_5[\psi]\,\psi \\ &&        + \alpha_1 
\log|\psi|^2 \psi
+ \alpha_2 (\arg\psi) \psi\,.
\end{array}
\end{equation}
$\F_5$ is invariant under the action of the group $\N$; i.e., 
$\overline{\F_5}=\F_5$.

Starting with the linear family $\F_0$, through iterated gauge 
generalizations and gauge closures with respect to the pure nonlinear 
gauge group $\N$, we have thus obtained a sequence 
\begin{equation}
\label{Fseq}
\F_0\subset\overline{\F_0} \subset \F_1 = \overline{\F_1} \subset 
\F_2 \subset \overline{\F_2} = \F_3 = \overline{\F_3} \subset \F_4 
\subset \overline{\F_4} = \F_5 \end{equation}
of families of nonlinear Schr\"odinger equations. 

The same procedure can be followed for the restricted gauge group 
$\N_0$. It turns out that there is an analoguous sequence of families 
$\R_j$ of NLSEs:
\begin{equation}
\label{Rseq}
\F_0\equiv\R_0\subset\overline{\R_0}^0 \subset \R_1 = 
\overline{\R_1}^0 \subset \R_2 \subset \overline{\R_2}^0 = \R_3 = 
\overline{\R_3}^0 \subset \R_4 \subset \overline{\R_4}^0 = \R_5\,. 
\end{equation}
The families $\R_j$ are subsets of the $\F_j$: \begin{equation}
\label{Rjfam}
\R_j = \F_j\restriction_{\nu_1(t)=\nu_1,\, \mu_0(t)=\mu_0,\,\mu_3(t) 
= -\nu_1,\,\alpha_2(t)= 0} \,.
\end{equation}
The only type of term that is not obtained in these families is the 
term $F_K$ (which is technically not well defined). Note furthermore, 
that here the parameters of the original linear family 
$\R_0\equiv\F_0$ remain invariant, $\np_1=\nu_1$ and $\mp_0=\mu_0$.

\section{Discussion of the Gauge-Generalized NLSE}\label{4} 
\subsection{Gauge-invariant parameters, Ehrenfest relations, and 
Galilei 
invariance}\label{4.1}
The group $\N$ transforms the family
$\F_5$ into itself. In fact, $N_{(\gamma,\Lambda)}$ acts (for all 
$t$) linearly on the eight gauge 
parameters$\v=\left(\nu_1,\ldots,\mu_5\right)$,
\begin{equation}
\label{para-trans1}
\left(\begin{array}{c}
\np_1\\ \np_2\\ \mp_0\\ \mp_1\\\mp_2\\ \mp_3\\\mp_4\\ \mp_5 
\end{array}\right) =
\left(\begin{array}{cccccccc}
\frac{1}{\Lambda} & 0&0&0&0&0&0&0 \\
-\frac{\gamma}{2\Lambda} & 1&0&0&0&0&0&0 \\ 0&0&\Lambda & 0&0&0&0&0 \\
-\frac{\gamma}{\Lambda} & 0&0&1&0&0&0&0 \\ \frac{\gamma^2}{2\Lambda} 
&-\gamma &0&-\frac{\gamma}{2}&\Lambda 
&0&0&0 \\
0&0&0&0&0&\frac{1}{\Lambda}&0&0 \\
0&0&0&0&0&-\frac{\gamma}{\Lambda}&1&0 \\ 
0&0&0&0&0&\frac{\gamma^2}{4\Lambda}&-\frac{\gamma}{2}&\Lambda 
\end{array} \right)
\left(\begin{array}{c}
\nu_1\\ \nu_2\\ \mu_0\\ \mu_1\\\mu_2\\ \mu_3\\\mu_4\\ \mu_5 
\end{array}\right)\,.
\end{equation}
One can show that the orbits of $\N$, for a fixed time $t$, are two 
dimensional on the space $\dot{\rz}^8_t := \left.\left\{\v \in 
\rz^{8}_t \right|\nu_1\neq 0\right\}$ and foliate $\dot{\rz}^{8}_t$ 
in two-dimensional leaves.
Hence there exist (in general, at least locally; but here in fact 
globally) six functionally independent parameters 
$\iota_0,\ldots,\iota_5$ invariant under the action of $\N$ 
\cite{N1995,DGN1995},
\begin{equation}
\label{gauge-inv1}
\begin{array}{c}\ds
\iota_0 = \nu_1\mu_0\,,\quad
\iota_1 = \nu_1\mu_2 -\nu_2\mu_1\,,\quad \iota_2 = 
\mu_1-2\nu_2\,,\quad
\iota_3 = 1 + \mu_3/\nu_1\,,\\ \ds
\iota_4 = \mu_4-\mu_1\mu_3/\nu_1\,,\quad \iota_5 = 
\nu_1(\mu_2+2\mu_5)-\nu_2(\mu_1+2\mu_4) 
+2\nu_2^2\mu_3/\nu_1\,.\\
\end{array}
\end{equation}
On the remaining two parameters $\a=(\alpha_1,\alpha_2)$, the 
transformation $N_{(\gamma,\Lambda)}$
acts as an affine transformation,
\begin{equation}
\left(\begin{array}{c}
\ap_1\\
\ap_2\end{array}\right) =
\left(\begin{array}{cc}
\Lambda & -\frac{\gamma}{2}\\
0       & 1
\end{array}\right)
\left(\begin{array}{c}
\alpha_1\\
\alpha_2\end{array}\right)
+
\left(\begin{array}{c}
\frac{1}{2}(\gamma\frac{\dot\Lambda}{\Lambda} 
-\dot\gamma)\\
- \frac{\dot\Lambda}{\Lambda}
\end{array}\right)\,.
\end{equation}
Thus there are two further independent parameters invariant under the 
action of $\N$ on the \emph{control space} $\dot\rz_t^{10}$ spanned 
by $\v$ and $\a$,
\begin{equation}
\label{gauge-inv2}
\iota_6 = \nu_1\alpha_1-\nu_2\alpha_2+\nu_2\frac{\dot\nu_1}{\nu_1} 
-\dot\nu_2 \,,\quad
\iota_7 = \alpha_2 - \frac{\dot\nu_1}{\nu_1}\,, \end{equation}
generalizing the result in Refs.~\cite{N1995,DGN1995} for the family 
of NLSEs derived in Refs.~\cite{DG1992,DG1994}. We call 
$\i=(\iota_0,\ldots,\iota_7)$ \emph{gauge-invariant 
parameters}. They are important for interpreting $\F_5$ and its 
subfamilies; for details, we refer to Ref.~\cite{DG1996}, where 
gauge-invariant parameters have been discussed in a slightly 
different context. 

The subfamilies $\F_j$ and $\R_j$, that are closed under the gauge 
groups $\N$ and $\N_0$, respectively, can now be characterized in 
terms of the vanishing of gauge-invariant parameters. Such a 
characterization is given in Tables 1 and 2, respectively. Note that 
$\nu_1$ and $\mu_3$ are themselves gauge-invariant parameters of the 
subfamilies $\R_j$. \begin{table}[t]
\begin{center}
\begin{tabular}{l@{\hspace*{4em}}r}
Table 1& Table 2\\
\begin{tabular}{c|cccccccc}
& $\iota_0$& $\iota_1$& $\iota_2$& $\iota_3$& $\iota_4$& $\iota_5$& 
$\iota_6$&
$\iota_7$\\
\hline
$\F_0$ & $\times$ & $\times$ & 0 & 0 & 0 & 0 & 0 & 0 \\ $\F_1$ & 
$\times$ & $\times$ & 0 & 0 & 0 & 0 & $\times$ & $\times$ 
\\
$\F_3$ & $\times$ & $\times$ & $\times$ & 0 & 0 & $\times$ & 
$\times$ & $\times$ \\
$\F_5$ & $\times$ & $\times$ & $\times$ & $\times$ & $\times$ & 
$\times$ & $\times$ & $\times$ \\
\end{tabular}
&
\begin{tabular}{c|cccccccc}
& $\iota_1$& $\iota_2$& $\iota_3$& $\iota_4$& $\iota_5$& $\iota_6$ 
&$\nu_1$& $\mu_0$ \\
\hline
$\R_0$ & $\times$ & 0 & 0 & 0 & 0 & 0 & $\times$ & $\times$ \\ $\R_1$ 
& $\times$ & 0 & 0 & 0 & 0 & $\times$ & $\times$ & $\times$ 
\\
$\R_3$ & $\times$ & $\times$ & 0 & 0 & $\times$ & $\times$ & 
$\times$ & $\times$ \\
$\R_5$ & $\times$ & $\times$ & $\times$ & $\times$ & $\times$ & 
$\times$& $\times$ & $\times$ \\
\end{tabular}
\end{tabular}
\end{center}
Table 1. Classification of subfamilies of $\F_5$ using 
gauge-invariants.

Table 2. Classification of subfamilies of $\R_5$ using 
gauge-invariants.
\end{table}

Some of these families show interesting behaviour. In the family 
$\F_1$ consider the time
dependence of the expectation values
$\langle\x\rangle_{\psi(t)}=\int_{\rz^3}\x\rho_t(\x)d^3x 
=\int_{\rz^3}\x\overline{\psi(\x,t)}\psi(\x,t) d^3x$. Then
\begin{eqnarray}
\label{Ehr-1}
\frac{d}{dt} \langle \x\rangle_{\psi(t)} &=& 
-2\nu_1\langle -i\grad \rangle_{\psi(t)}\,, \\ \label{Ehr-2}
{\frac{d^2}{dt^2}} \langle \x\rangle_{\psi(t)} &=& -2\iota_0 \langle 
-\mbox{grad} V \rangle_{\psi(t)} 
+\iota_7 \frac{d}{dt} \langle \x\rangle_{\psi(t)} \,; \end{eqnarray}
i.e., we have the analogue of the first and second Ehrenfest 
relations for $\F_1$.
The center of a
non-stationary solution behaves like a classical system under a 
conservative force and a frictional force proportional to the 
velocity. For the linearizable subfamily 
$\overline{\F_0}\subset\F_1$, the frictional term disappears 
($\iota_7=0$). This is plausible: a linear or nonlinear 
quantum-mechanical evolution equation and its $\N$-transform describe 
physically equivalent systems \cite{DG1996}.

The first Ehrenfest relation \Ref{Ehr-1} holds for all members of 
$\F_5$.
This shows that the physical systems described by $\F_5$ have 
something in common. For $\F_2,\ldots,\F_5$ there are additional 
terms in the second Ehrenfest relation \Ref{Ehr-2}, which are 
connected with the quantum-mechanical diffusion current 
\cite{DG1992,DG1994}.

The free linear SE ($V\equiv 0$) is invariant under the centrally 
extended Galilei group $G_e(3)$ including time translations. Consider 
the $G_e(3)$ invariance of $\F_5$ and its subfamilies. $\F_5$ is 
invariant under
$T(t)$, if the gauge-invariant parameters $\i$ are time independent. 
If in addition $\iota_3=\iota_4=\iota_7=0$ the equations are 
invariant under $G_e(3)$. The generator of time translations is 
represented via a nonlinear operator $H_{nl}$, $i\dt\psi = 
H_{nl}[\psi]$, as in Eqs.~\Ref{NLSE}--\Ref{NLSE2}, while all other 
generators of $G_e(3)$ are as usual represented linearly. Hence, one 
has a nonlinear representation of $G_e(3)$ (see also Refs. 
\cite{N1995,RW1993,DG1993a}). 

\subsection{Gauge-generalized NLSE as a unification}\label{4.2} Now 
we are ready to understand the connection between various proposals 
for nonlinear terms to be added to the linear Schr\"odinger equation. 
Such terms have often been chosen in a physically guided, but 
\emph{ad hoc} way.
Some proposed terms have been based directly on fundamental 
considerations. Our attempt is of the latter type. Its foundation, 
the physical equivalence of theories and the resulting group of 
nonlinear gauge transformations (together with gauge generalization 
and gauge closure) reflects some of the structure of quantum 
mechanics. Consequently the
family $\F_5$ exhibits a common, fundamental basis for some of the 
proposed NLSEs. Let us consider some of the particular nonlinearities 
that have been proposed.

\subsubsection{Logarithmic nonlinearity} Based on the observation 
that all linear evolution equations for physical
quantities are approximations of nonlinear evolutions (except for the 
Schr\"odinger equation) Bialynicki-Birula and Mycielski \cite{BM1976} 
added a
(local) nonlinear term $F(|\psi|^2)$. They used the separation 
property 
to show that $F$ has to be logarithmic,
$F(|\psi|^2)=-b\ln|\psi|^2$. Their NLSE (the BM-family) is 
\begin{equation}
\label{BBM}
i\hbar\dt\psi = \left(-\frac{\hbar^2}{2m}\Delta+ V - b 
\ln|\psi|^2\right) \psi \,.
\end{equation}
This NLSE is contained in $\F_5$ with
\begin{equation}
\begin{array}{c}
\ds
\nu_1=\frac{\hbar}{2m},\quad
\mu_2=\frac{\hbar}{4m},\quad
\mu_3=-\frac{\hbar}{2m},\quad
\mu_5=-\frac{\hbar}{8m},\quad
\mu_0=\frac{1}{\hbar},\quad
\alpha_1 = -\frac{b}{\hbar},
\end{array}
\end{equation}
and the other coefficients vanishing. Note that in order to obtain 
this logarithmic term in our gauge generalization, we had to allow 
for a time-dependent group parameter $\gamma=\gamma(t)$. 

\subsubsection{Nonlinearity proportional to the phase} One of many 
examples of a heuristic implementation of dissipation in quantum 
mechanics is the approach by Kostin \cite{Kos1972}. Starting with a 
frictional term proportional to the expectation of the momentum 
operator in the (second) Ehrenfest relation, Kostin motivated adding 
a nonlinear term proportional to the phase of $\psi$ to the linear 
Schr\"odinger equation; i.e. (with $f\in\rz$), %
\begin{equation}
\label{K}
i\hbar\dt\psi = \left(-\frac{\hbar^2}{2m}\Delta+ V +\frac{\hbar f}{m} 
\arg\psi \right) \psi \,. \end{equation}
Kostin's NLSE (the K-family) is contained in $\F_5$ with %
\begin{equation}
\begin{array}{c}
\ds
\nu_1=\frac{\hbar}{2m},\quad
\mu_2=\frac{\hbar}{4m},\quad
\mu_3=-\frac{\hbar}{2m},\quad
\mu_3=-\frac{\hbar}{8m},\quad
\mu_0=\frac{1}{\hbar},\quad
\alpha_2 = \frac{f}{m},
\end{array}
\end{equation}
and the other coefficients vanishing. To obtain this term in our 
approach, we had to assume that $\Lambda=\Lambda(t)$ can be a 
function of time. Obviously, $\arg\psi$ is not well
defined; this is reflected in the problem of gauge transformations 
with $\Lambda \neq \pm 1$, discussed in Section~\ref{2.4}.

\subsubsection{Nonlinearity from diffeomorphism group 
representations} The approach of Doebner and Goldin 
\cite{DG1992,DG1994} is motivated by fundamental considerations. The 
generic kinematical symmetry algebra $S(\rz^3)$ on $\rz^3$ is a 
semidirect sum of the Lie algebra of real smooth functions $f\in 
C^\infty(\rz^3)$, and the Lie algebra of vector fields $X\in 
\X(\rz^3)$, or equivalently a local current algebra on $\rz^3$ 
\cite{DS1968,GS1970,G1971}. $\X(\rz^3)$ is the Lie
algebra of a subgroup of the group of
diffeomorphisms of $\rz^3$ (diffeomorphisms trivial at infinity). The 
functions $f\in C^\infty(\rz^3)$ can be interpreted physically as 
classical position observables and the vector fields $X\in\X(\rz^3)$ 
as classical kinematical momenta. Then a quantization map $\mathcal 
Q$ represents the kinematical algebra $S(\rz^3)$ by self-adjoint 
operators in the single particle Hilbert space $\H^{(1)}$.
Under physically motivated assumptions, all such representations 
$\mathcal Q$ can be classified up to unitary equivalence by a real 
parameter $D$ with the dimensionality of a diffusion coefficient 
[length$^2$/time]. The presence of such a family of inequivalent 
representations reflects the richness of $\X(\rz^3)$. The method can 
be
generalized to any smooth manifold \cite{ADT1983}. 

To obtain some information about the evolution equation of $\psi$, 
local probability
conservation (for pure states) is assumed \cite{DG1992}, or a 
generalized first Ehrenfest
relation is postulated \cite{DH1995,DN1996}. Then the time-dependent 
probability density and current are related through an equation of 
Fokker-Planck type,
\begin{equation}
\label{fpe}
\dt \rho = -\frac{\hbar}{m}\nabla\cdot\J + D\Delta\rho\,.
\end{equation}
This restricts the evolution equation of $\psi$ to the form 
\begin{equation}
\label{nse}
i\hbar\dt\psi = \left(-\frac{\hbar^2}{2m} \Delta + V\right) \psi 
+ i\,\frac{\hbar D}{2}
\frac{\Delta\rho}{\rho}\,\psi
+ R[\psi]\,\psi\,,
\end{equation}
where $R[\psi]$ is an arbitrary real-valued (nonlinear) operator. The 
form of the pure imaginary functional, $\Delta\rho / \rho\,$, is 
enforced. If $R[\psi]$ is assumed to be of a similar form, i.e., if 
it is (i) complex homogeneous
of degree zero, (ii) a rational function with derivatives of no more 
than second order occuring only in the numerator, and (iii) invariant 
under the 3-dimensional Euclidean group $E(3)$, then a five parameter 
family of NLSEs (the DG-family) is obtained: \begin{equation}
\label{Rf}
R[\psi] = \hbar D' \sum_{j=1}^5 c_j R_j[\psi] \,, \end{equation}
with the $R_j$ as in Eq.~\Ref{Rj}.
Obviously this is a special case of $\F_5$, where 
$\alpha_1=\alpha_2=0$ and all gauge parameters are time-independent:
\begin{equation}
\label{DGpara}
\begin{array}{c}
\ds
\nu_1 = -\frac{\hbar}{2m}\,,\quad
\nu_2 = \frac{\hbar D}{2}\,,\quad
\mu_0 = \frac{1}{\hbar}\,,\quad
\mu_1 = \hbar D' c_1\,,\quad
\mu_2 = \hbar D' c_2 - \frac{\hbar}{4m}\,,\\[4pt] \ds
\mu_3 = \hbar D' c_3 + \frac{\hbar}{2m}\,,\quad \mu_4 = \hbar D' 
c_4\,,\quad
\mu_5 = \hbar D' c_5 + \frac{\hbar}{8m}\,,\quad \alpha_1 = \alpha_2 = 
0\,.
\end{array}
\end{equation}
The equation proposed by Guerra and Pusterla in connection with de 
Broglie's double solution theory \cite{GP1982} is contained in this 
family, with $D=0$, $c_1=c_3=c_4=0$, $c_5=-\frac{1}{2}c_2$. 

\section{Summary}\label{5}

To summarize, we have taken a small step toward a nonlinear quantum 
theory which could be physically relevant, by discussing nonlinear 
evolution equations derived from fundamental considerations.

Under the assumption that all measurements are positional 
measurements performed at different times, we derived a group of 
nonlinear gauge transformations $\G$, including the usual linear 
ones. Applying these transformations to a linear Schr\"odinger 
equation, we obtained nonlinear ones, and after gauge generalization 
and gauge closure we reached a family $\F_5$ of nonlinear 
Schr\"odinger equations. Certain subfamilies of $\F_5$ were motivated 
originally by different physical ideas and different mathematical 
structures. Thus $\F_5$ is a \emph{unification} of these NLSEs: the 
BM-family, the K-family, and the DG-family. It is surprising, and 
also satisfying, when different structures and lines of reasoning 
yield the same or compatible results.
This is an indication that these structures have a common origin. If 
there is some deeper reason for this, beyond the gauge generalization 
process described here, we have not yet unveiled it.

Moreover, our discussion may show how to circumvent some formal 
arguments against nonlinear quantum theory put forth by Gisin and 
others \cite{G1990,P1991,Gi1995}; in connection with nonlocal 
effects, we refer especially to \cite{LN1997}. We have not touched on 
other problems of nonlinear quantum theory, such as the concept of 
mixed states (see Ref.~\cite{N1997}), or discussed the physical 
interpretation of a (necessarily non-selfadjoint) nonlinear 
Hamiltonian.

\subsection*{Acknowledgements}
We would like to thank W.~L\"ucke for fruitful discussions on the
topic of this paper. 
G.A.G. acknowledges hospitality from the Arnold Sommerfeld
Institute for Mathematical Physics, Technical University of
Clausthal, and travel support from the DAAD (Deutscher Akademischer
Austauschdienst) and Rutgers University.
H.D.D. acknowledges the really essential help of Dr. G. A. Oswald FRCP 
and the Coronary Care Unit, Princess Elizabeth Hospital, Guernsey and 
of Prof. Dr. Kreuzer and Dr. Schulze, Unit 1023 ``HIS'', University 
Hospital G\"ottingen, where an earlier draft of this paper was 
written. 
\end{document}